\documentclass[onecollarge,natbib]{svjour2}
\bibpunct{[}{]}{;}{n}{}{,} % to get "[numbered]" references from natbib
\smartqed  % flush right qed marks, e.g. at end of proof
\usepackage{graphicx}
\usepackage{comment}
\usepackage{epsfig}
\usepackage{wrapfig}
\usepackage{lipsum}

\newcommand{\bea}{\begin{eqnarray}}
\newcommand{\eea}{\end{eqnarray}}
\newcommand{\be}{\begin{equation}}
\newcommand{\ee}{\end{equation}}

\newcommand{\lqcd}{\Lambda_\mathrm{QCD}}
\def\dpdf#1{D_{#1}}

\newlength\savedwidth

\journalname{Few-Body Systems}
\begin{document}

\title{Double parton scattering and 3D proton structure:
A Light-Front analysis
}


\author{Matteo Rinaldi         \and
        Sergio Scopetta \and Marco Traini \and Vicente Vento %etc.
}


\institute{M. Rinaldi \at
              Dipartimento di Fisica e Geologia,
Universit\`a degli Studi di Perugia, Via A. Pascoli, I-06123 Italy \\
INFN section of Perugia
              \email{matteo.rinaldi@pg.infn.it}           %  \\
           \and
           S. Scopetta \at
 Dipartimento di Fisica e Geologia,
Universit\`a degli Studi di Perugia, Via A. Pascoli, I-06123 Italy \\
INFN section of Perugia
\and
M. Traini \at
Dipartimento di Fisica, Universit\`a degli Studi di Trento,
Via Sommarive 14, I - 38123 Povo (Trento), Italy \\ 
INFN - TIFPA, Trento, Italy
\and
V. Vento \at
Departament de Fisica Te\`orica, Universitat de Val\`encia
and Institut de Fisica Corpuscular, Consejo Superior de Investigaciones
Cient\'{\i}ficas, 46100 Burjassot (Val\`encia), Spain              
}

\date{Received: date / Accepted: date}

\maketitle

\begin{abstract}
Double parton scattering, occurring in high energy hadron-hadron collisions,
e.g. at the LHC, is usually investigated through model dependent analyses of the so
called effective cross section $\sigma_{eff}$. We present a dynamic approach to
this fundamental quantity making use of  a
Light-Front model treatment.  Within such a framework
$\sigma_{eff}$ is initially evaluated at low energy scale { using} the
model and then, through QCD evolution, at high energy scale to reach the
experimental conditions. Our numerical outcomes are consistent with the
present experimental analyses of data in
kinematical region we investigate. An important result of the present
work is the $x_i$ dependence of $\sigma_{eff}$, a feature directly connected to
double parton correlations and which could unveil new information on the
three dimensional structure of the proton.
\end{abstract}

\section{Introduction}

In hadron-hadron collisions, possible signals of Multi Parton Interactions
(MPI), due to the interactions of at least two partons of each hadron, have
been  investigated  { for a long time}  \cite{paver}. Recently, thanks to
the the
possibilities offered by the Large Hadron Collider (LHC), RUN 2, a lot of
attention on this subject has been paid  \cite{gaunt,diehl_1,manohar_1}.
In the present talk, we focus our scrutiny on the simplest MPI process,
double parton scattering (DPS). { The 
cross section for this process}, assuming
factorization, depends on non-perturbative quantities, the so called double
parton distribution functions (dPDFs), which describe the joint probability
of finding two partons at a given transverse distance with given longitudinal
momentum fractions. dPDFs are therefore related to the three-dimensional (3D) 
nucleon  structure \cite{calucci} and are non-perturbative 
quantities.
{ Tractable} model calculations of the dPDFs have
been proposed in the last few years (e.g. Refs. \cite{manohar_2,noi1,noi2}).
In particular, 

%\newpage

\noindent in Refs. \cite{noi2,lc14}, dPDFs have been
evaluated by means of a Light-Front (LF) approach, which allows { one} to
achieve a  Poincar\'e covariant treatment and then to reproduce
the essential  dPDFs sum rules.
Let us stress that even if there are no clear data from which dPDFs can be
extracted, a signature of DPS has been observed and measured in several 
experiments \cite{afs,data0,data2,data3,data4,data5}. In particular
the DPS cross section is estimated through  
the so called  ``effective cross section'', $\sigma_{eff}$ and
evaluated in model dependent ways. Despite  large
error bars $\sigma_{eff}$ is found  to be constant  w.r.t. the
center-of-mass energy 
of the collision and the longitudinal momentum fractions carried by the
interacting partons.
In order to calculate $\sigma_{eff}$ in our dynamical model, a useful expression 
has been derived in Ref. \cite{lettern}. Thanks to that study,  $\sigma_{eff}$ can be
be directly evaluated from the calculations of dPDFs. In particular, in
\cite{lettern} and here, use has been made of the results discussed in Ref.
\cite{noi2}, in which dPDFs have been estimated by means of a LF model.
In the next sections the main results of the calculations of  $\sigma_{eff}$ will be 
properly shown and described.

\section{The effective cross section}

In this section, starting from the phenomenological definition of
$\sigma_{eff}$, a direct relation between this quantity and standard parton
distribution functions (PDFs) and dPDFs will be shown. $\sigma_{eff}$, is
usually defined through a ``pocket formula'' when a DPS
process produces 
two final states $A$ and $B$  \cite{diehl_1}:

\begin{eqnarray}
\label{pocket}
\sigma_{eff} = { m \over 2 }
{ \sigma_A^{pp'} \sigma_B^{pp'} \over 
\sigma^{pp}_{double}
}~.
\end{eqnarray}
Here,  $m= 1$
if $A$ and $B$ are identical and $m=2$ if they are different.
In the numerator,
$\sigma_{A(B)}^{pp'}$ is the differential
cross section for the inclusive process $pp' \rightarrow A (B) + X$ and
reads as follows:

\begin{eqnarray}
\label{s_single}
\sigma_{A(B)}^{pp'}(x_{1(2)},x'_{1(2)},\mu_{1(2)}) & = &
\sum_{i,k}  F_{i}^{p} (x_{1(2)},\mu_{1(2)})
F_{k}^{p'} (x'_{1(2)},\mu_{1(2)}) \,
\hat \sigma_{ik}^{A(B)}(x_{1(2)},x'_{1(2)},\mu_{1(2)})~, 
\end{eqnarray}
where here $F_{i(k)}^p$ is the standard single PDF, the indices read
 $i,k = \{q, \bar q, g \}$ and $\mu_{1(2)}$ is
the 
factorization
scale for the  process $A (B)$.
On the other hand side, the denominator of Eq. (\ref{pocket}) is the
differential double parton scattering cross section:

\begin{eqnarray}
\nonumber
\sigma^{pp}_{double}(x_1,x_1',x_2,x_2',\mu_1,\mu_2) &=&
{m \over 2} 
\sum_{i,j,k,l} \int
D_{ij}(x_1,x_2; {\bf {k_\perp}},\mu_1,\mu_2)\,
\hat \sigma_{ik}^A(x_1,x_1',{\mathbf{\mu_1}})\,
\\
&\times&
D_{kl}(x_1',x_2'; {\bf {- k_\perp}},\mu_1,\mu_2) 
\hat \sigma_{jl}^B(x_2,x_2',{\mathbf{\mu_2}})\,
{d {\bf {k_\perp}} \over (2 \pi)^2}~.
\label{4}
\end{eqnarray}
In Eq. (\ref{4}), $D_{ij}(x_1,x_2; {\bf {k_\perp}})$,  often called 
``double generalized parton distributions'' ($_2GPDs$,
\cite{blok_1,blok_2}), depends on ${\bf {k_\perp}}$, the transverse
momentum
{ imbalance} of the partons 1 and 2. $_2GPDs$
 { are} the Fourier transform of the dPDFs 
$D_{ij}(x_1,x_2; {\bf {r_\perp}})$, which depend on ${\bf {r_\perp}}$, the 
transverse separation, in coordinate space, between two
partons. $D_{ij}(x_1,x_2; {\bf {r_\perp}})$ represents the
probability of finding  parton pairs $i,j$ with longitudinal momentum 
fractions $x_1,x_2$ and transverse distance ${\bf {r_\perp}}$. This is the 
non-perturbative input describing soft-physics.
Let us point out the physical content of  $\sigma_{eff}$.
In Eq. (\ref{pocket}), if the occurrence of the  $B$ were not biased
 by that of the $A$, instead of the ratio
$\sigma_B/\sigma_{eff}$ one would read
$\sigma_B/\sigma_{inel}$, representing the probability
to have  $B$ once $A$ has taken place assuming rare
hard multiple collisions. The difference between $\sigma_{eff}$ 
and $\sigma_{inel}$ 
measures therefore correlations between the interacting partons
in the colliding proton.

%\begin{eqnarray}
%\hat \sigma_{ij} (x,x') = C_{ij} \bar \sigma(x,x')~,
%\label{cij}
%\end{eqnarray}

%\begin{eqnarray}
%\label{color}
%C_{gg}:C_{qg}:C_{qq}=1:(4/9):(4/9)^2~.
%\end{eqnarray}

Now  a suitable expression of $\sigma_{eff}$ for
microscopic
calculations, in particular highlighting the   $x_i$-dependence of
$\sigma_{eff}$, will be derived.
First of all, let us assume that  heavy quark flavors do not contribute in
the process and 
the { that} elementary cross section, appearing in Eqs. (\ref{s_single})
and
(\ref{4}), is basically (see Ref. \cite{lettern} for details) $\hat
\sigma_{ij} (x,x') = C_{ij} \bar \sigma(x,x')$,
where $\bar \sigma(x,x')$ is a universal function, and $C_{ij}$
are color factors, with $i,j=q,\bar q,g$,  which stay in the ratio
$C_{gg}:C_{qg}:C_{qq}=1:(4/9):(4/9)^2$. 

At this point, inserting  Eqs. (\ref{s_single})-(\ref{4}) in Eq.
(\ref{pocket}) together with these assumptions, one finds the following
formal
expression of $\sigma_{eff}$:
\begin{eqnarray}
\label{simple}
\sigma_{eff} (x_1,x_1',x_2,x_2') = 
{
\sum_{i,k,j,l }  F_{i} (x_{1})
F_{k} (x_{1}')
F_{j} (x_{2})
F_{l} (x_{2}')
C_{ik}
C_{jl}
\over
\sum_{i,j,k,l} 
C_{ik}
C_{jl}
\int
D_{ij}(x_1,x_2; {\bf {k_\perp}})
D_{kl}(x_1',x_2'; {\bf {-k_\perp}}) 
{d {\bf {k_\perp}} \over (2 \pi)^2}
}
~.
\end{eqnarray}

\begin{wrapfigure}{l}{0.3\textwidth} 
 \includegraphics[width=5cm]{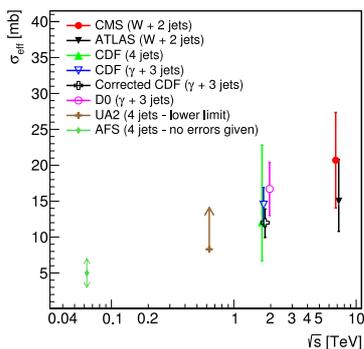}
 \caption{\scriptsize{Centre-of-mass energy dependence of $ \sigma_{eff}$
measured by different experiments using
different processes
\cite{afs,data0,data2,data3,data4,data5}.
The figure is taken from \cite{data5}.}}
\label{fig:Fig1}
 \vspace{15pt}
\end{wrapfigure}

As one can see, in the above equation, assuming a complete factorized
ansatz for dPDFs in terms of standard PDFs, e.g., $D_{ij}(x_1,x_2; {\bf
{k_\perp}}) \propto F_{i} (x_{1})F_{j} (x_{2})T({\bf {k_\perp}})  $, one
obtains a constant value of $\sigma_{eff}$ w.r.t. to $x_i$.
This scenario is realized neglecting  double parton correlations.
This is the kind
of condition often guessed for the extraction of $\sigma_{eff}$ from data
since there is no experimental information on dPDFs.
The problems related to the uncorrelated ansatz are discussed in a 
number of papers
(see, e.g., Ref.~\cite{gaunt,diehl_1,muld}). 
In particular, in the valence region we are interested here,
this assumption
is not supported by model calculations \cite{manohar_2,noi1,noi2}, due to
the presence of double parton correlations.
Before describing our results,
the present experimental scenario is
shown   in Fig. \ref{fig:Fig1}, where $\sigma_{eff}$, measured by  
different experiments \cite{afs,data0,data2,data3,data4,data5},
which analyzed the latter quantity 
with different final states at different values 
of the center-of-mass
energy, $\sqrt{s}$,
has been
plotted. 
As one can see, a constant value of $\sigma_{eff}$ 
 could be consistent with the data within the large experimental errorbars,
which could hide the fundamental information on the possible
$x_i$-dependence. Let us stress that such important dependence could open
new
ways to access the 3D nucleon structure \cite{calucci}.

Nowadays, 
fundamental information on the 3D structure of the nucleon, related to
the transverse position
of partons,  is usually
investigated thorough the study and the measurements of hard-exclusive
processes  such as  deeply virtual Compton scattering
(DVCS), which allows to access the  Generalized Parton Distributions (GPDs)
(see Ref. \cite{gui} for recent developments).
{ Further information could be obtained  via the investigation of} dPDFs
in
its full $x_i$
dependence  which can be extracted from  $\sigma_{eff}$ in DPS.
Let us remark that the information encoded in dPDFs are anyhow complementary
to those provided by GPDs in impact parameter space. In fact, the latter
quantities are one-body densities describing the probability of finding a
parton with longitudinal momentum fraction  $x$ at a given transverse
distance from the center of the target, while, dPDFs are
two-body distributions related to the {\it relative} distance between two
partons with  given $x_1$ and $x_2$. In other words, the measurements of
DPS could access information on the  
average transverse distance of two fast or slow partons: 
a very interesting dynamical feature, not included in GPDs.

\section{{\bf Calculation of the effective cross
section within a LF approach}}

dPDFs can not be easily evaluated  within
QCD while they can be estimated at a low
energy scale, $Q_0 \sim \lqcd$, using models, 
as extensively done for the  PDFs (e.g. Ref. \cite{trvlc}).
The results of these analyses should then be evolved using perturbative
QCD (pQCD) in order to match data taken at an energy scale $Q>Q_0$. 
dPDF evolution is
discussed  e.g. in  
Refs.~\cite{gaunt,diehl_1,Shelest:1982dg,ruiz, 1x2}.
The analysis we are presenting makes use of
the Poincar\'e covariant Light-Front model approach,  allowing for  a
correct starting point for a precise pQCD evolution.
Thanks to this feature, our model calculations can be relevant for the
analysis of high-energy data.

The model, chosen in order to grasp the most relevant
features of dPDFs, is the one presented and described in  Ref.
\cite{LF2}, being already used for the studies of other different
distributions (see, e.g.,

\begin{figure}[t]
\label{fig:Fig2}
\begin{center}
\epsfig{file=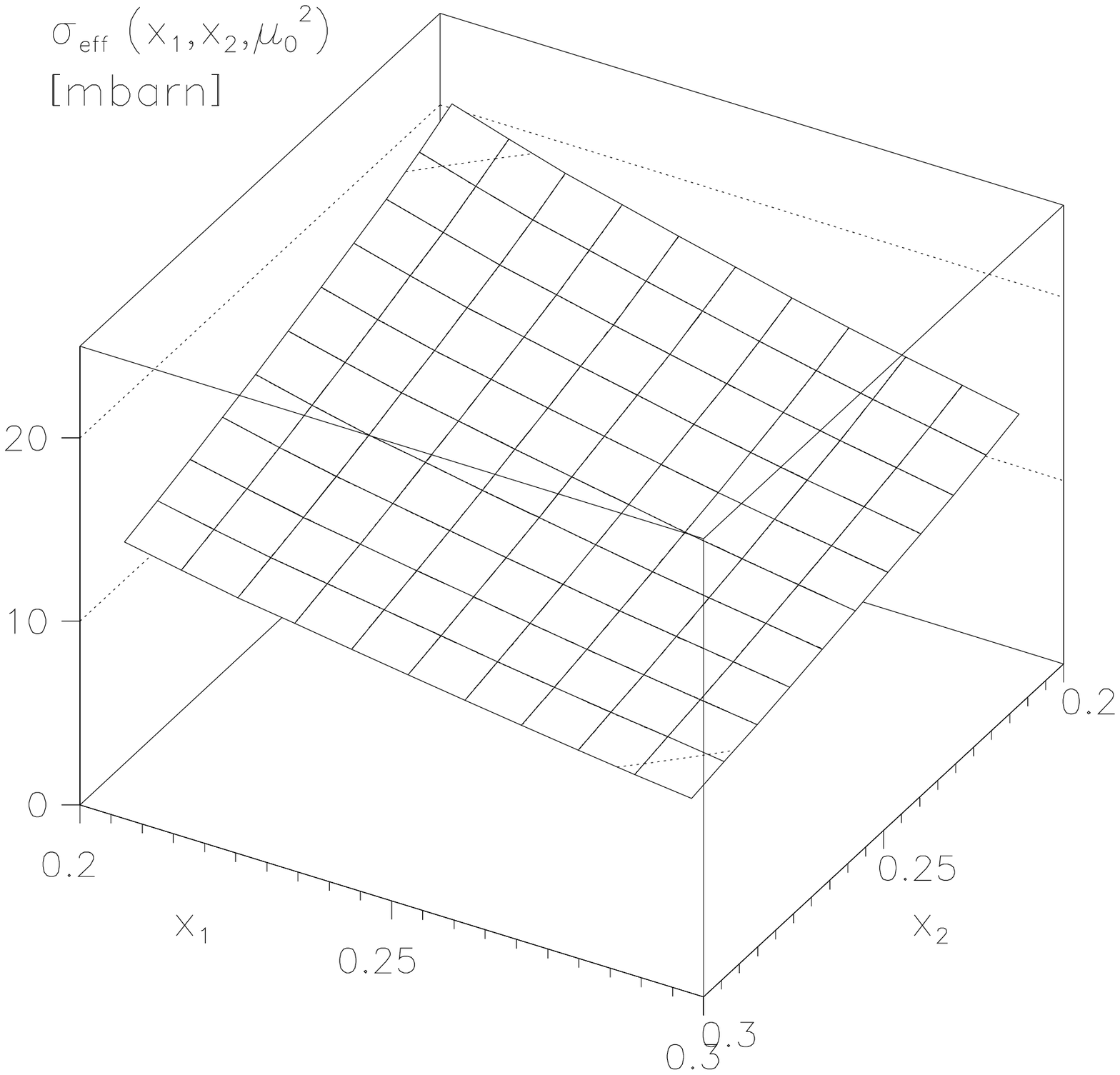,
%hoffset=-40 voffset=-240 hscale= 60 vscale=60, 
width=5.cm
%\columnwidth
}
%\special{psfile=Grafico2.eps, 
\epsfig{file=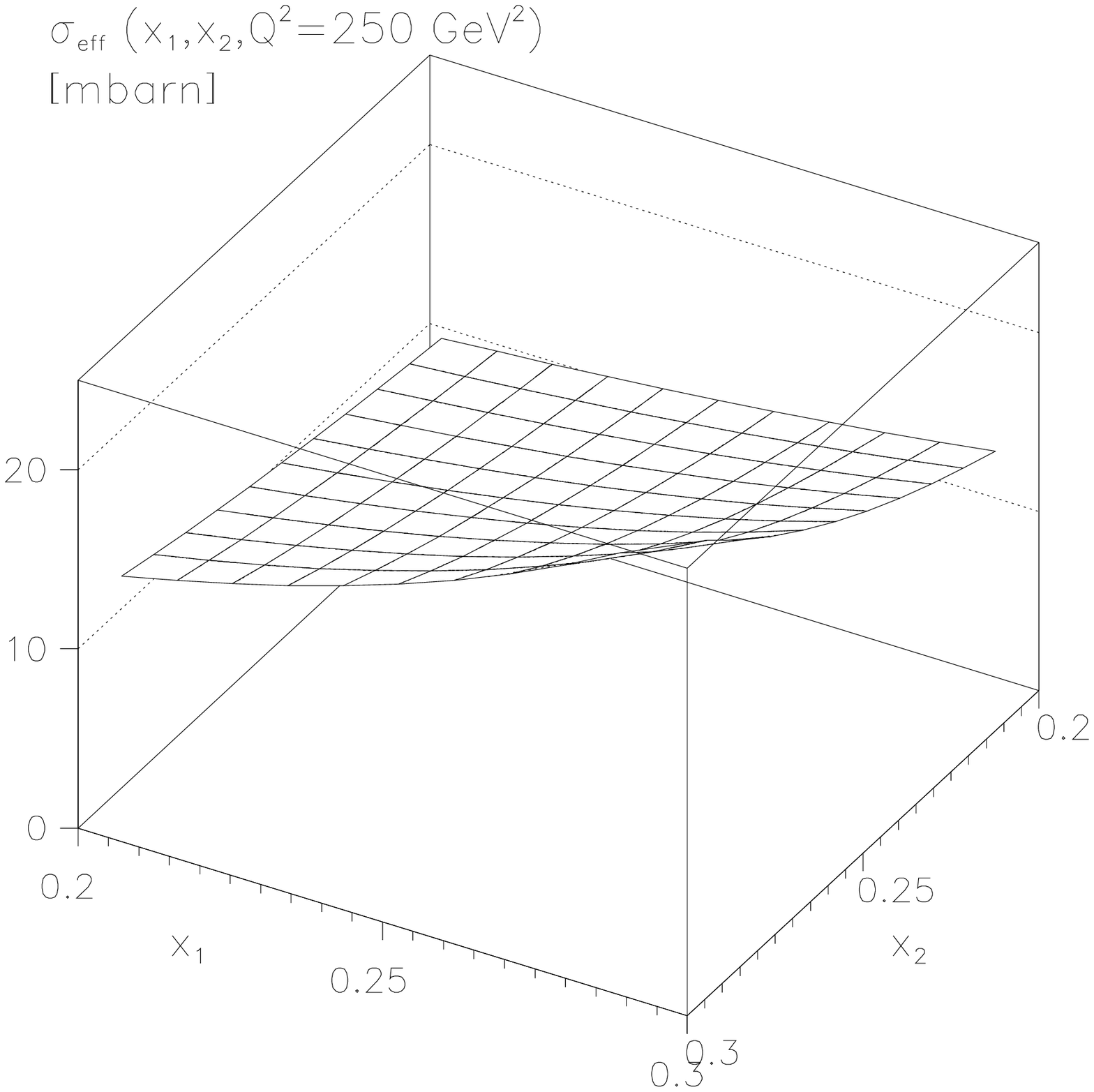, 
%hoffset=220 voffset=-240 hscale= 60 vscale=60,
width=5.cm
%\columnwidth
}
\end{center}
\caption{$ \sigma_{eff}(x_1,x_2,Q^2)$ for the values of $x_1, x_2$
measured in Ref. \cite{afs}. Left panel:
hadronic scale; right panel: $Q^2$ = 250 GeV$^2$.}
\end{figure}

\noindent Refs. \cite{pasquini,marco2} and references
therein). The explicit  calculation of 
$\dpdf{ij}(x_1,x_2;{\bf k_\perp},\mu)$
has been described in details
in Refs.~\cite{noi2,lc14}.  In particular, let us point out that
strong correlation effects are present  at the scale of the model and still
sizable, in the valence region, at the experimental scale, i.e. after pQCD
evolution. At  low values of $x$, presently studied at the LHC, 
correlations become less relevant, 
although their effects are still important for the spin-dependent
contributions  to unpolarized proton scattering.
In order to evaluate  $\sigma_{eff}$, the single and the double PDFs,
appearing in Eq. (\ref{simple}) have been calculated imposing  a unique 
factorization scale $\mu_1 = \mu_2 = \mu_0$ (as in, e.g., Refs.
\cite{gaunt,manohar_1,Diehl:2014vaa}), where $\mu_0$ is
the hadronic scale, where only valence quarks  are present.
Here and in Ref. \cite{lettern}  we focus our analysis on the kinematics 
of the old AFS data \cite{afs}, i.e. 
$x_1 \simeq x_1', x_2 \simeq x_2'$ 
and  $0.2 \leq x_{1,2} \leq 0.3$ and 
the average energy scale turns out to 
be $Q^2 \simeq 250$ GeV$^2$.
The results of the calculations are shown in Fig. 2, at the
scale of the model, $\mu_0^2 \simeq 0.1$ GeV$^2$, and after non-singlet
evolution to $Q^2$ (details on the fixing of the  hadronic scale and on the
calculation of the QCD evolution can be found in Ref. \cite{noi2}).
As one can see, a strong $x_{1,2}$ dependence has been found in this narrow
kinematical range at both the experimental and  the model scales.
One should notice that the three old experimental extractions of
$\sigma_{eff}$ 
\cite{afs,data0,data2}, which include the valence region
(cf. Fig. 1), lie
in the obtained range of values of $\sigma_{eff}$ shown in Fig. 2.
Let us remark two important issues on the pQCD evolution of
dPDFs. In our approach, there is only evolution on $x_{1,2}$, being 
the one on $k_\perp$ still an open challenge and, furthermore, since we 
are interested, for the moment being, on the valence
region, only the homogenous part of the evolution equations has been taken
into account in our calculation. 
Moreover, in Ref. \cite{lettern}, the average value of $\sigma_{eff}$, $\bar
\sigma_{eff} \sim 10.9$ mb, is in good agreement with the
experimental data reported in Fig. 1.

\section{Conclusions}

In the present work, a suitable 
expression of
$\sigma_{eff}$ for microscopic
calculations, in terms of dPDFs and standard PDFs, has been derived.
In order to perform the calculation of $\sigma_{eff}$, use has been made of
a relativistic Poincar\'e covariant quark model.
In particular, our investigation predicts
an  $x_i$ dependence of $\sigma_{eff}$, whose values  are consistent with
data including the valence region. Moreover, such dependence, found when a
non singlet evolution
of the valence distributions is performed, as well as
when perturbatively generated gluons are included 
into the scheme,  is a
feature not 
easily read in the available data.
The measurement of $\sigma_{eff}$ in restricted $x_i$ ranges
would lead, therefore, to a first signature of 
double parton correlations in the proton,  a novel
and interesting
aspect of the 3D structure of the nucleon.

\vskip 1cm

\noindent
This work was supported partially through the McCartor Fund
Fellowship program under the auspices of International Light-Cone Advisory
Committee
(ILCAC).

\end{document}